\documentclass[10pt]{article}
\author{Dmitri Noshchenko$^{1,2}$ \\ \ \\ \small{1. Institute of Cosmophysical Researches and Radio Wave Propagation FEB RAS}
\\ \small{2. Vitus Bering Kamchatka State University} \\ \ \\ \hline \small{\texttt{d951039@gmail.com}}}
\title{Bilinear gauge operators}
\usepackage{mystyle}

\begin{document}
\maketitle

\begin{abstract}
We construct a family of bilinear differential operators which
satisfy certain gauge properties. These operators can be naturally
associated with $q$-deformations of classical integrable hierarchies.
In particular, we consider the case when gauge function $=$ Hurwtiz-type partition function.
\end{abstract}

\section{Introduction}

Our aim is to represent bilinear Hirota-type differential operators of the form
\begin{equation}\label{Ddef}
D^n=\sum_{0<i,j\leq n}C_{i,j}(q)\frac{\partial^i}{\partial x^i}\frac{\partial^j}{\partial {x'}^j}_{x=x'}, \quad q \in \mathbb{C},
\end{equation}
which satisfy gauge condition for some fixed $g$:
\begin{equation}\label{gp}
D\left\langle gf_1,gf_2\right\rangle=g^{2+s}D\left\langle f_1,f_2\right\rangle
\end{equation}
where $f_1(x),f_2(x')$ are smooth complex-valued functions, $s \in \mathbb{Z}$ and $g$ has perturbative expansion
in powers of $\hbar=\log(q)$:
\begin{equation}
g(x,e^{\hbar})=\sum_{0\leq i<\infty} S_i(x) \hbar^i
\end{equation}
Let $S_0(x)=e^x$, so when $\hbar=0$, $D^n$ coincides with the ordinary Hirota derivative \cite{1}
(it is uniquely defined using gauge condition).
We claim that such operators can be used as building blocks for some interesting equations
(relations with $q$-Schur functions are mostly conjectured, see \cite{2,5} for more reference).

Of course, it can be generalized to multi-linear multi-dimensional case (although there may me some obstacles,
and we'll show this later).
``Good'' choice of gauge function $g$ (a proper sum of $q$-exponentials) can lead to non-trivial properties.
One of our goals is to understand the structure of bilinear operators of the type
\begin{equation}
P^m(D_{(g)}^1,D_{(g)}^2,\dots)=0
\end{equation}
where $P$ is a quasi-homogeneous polynomial (monomials are labeled by partitions of $m$) with $q$-deformed coefficients.
We add $(g)$ as lower index to emphasise dependence on a gauge.

\section{$q$-exponential case}
Before producing the general construction, consider the following example. Choose $g(x,q)$ as the $q$-exponential:
\begin{equation}\label{q-exp}
e_q^x=\sum\limits_{k=0}^{\infty}\frac{(1-q)^k}{(q;q)_k}x^k,
\end{equation}
where $(a,q)_k$ is a $q$-shifted factorial
\begin{equation}
(a,q)_k=\prod\limits_{s=0}^{k+1}(1-aq^s), k=1,2,\dots
\end{equation}
Recall definitoin of $q$-derivative:
\begin{equation}
d_q(f)=\frac{f(x)-f(qx)}{(q-1)x}
\end{equation}
Consider the following $q$-difference operator:
\begin{equation}\label{Dq}
\Delta_q^1(f_1(x),f_2({x'}))={\frac { \left( -f_1 \left( x \right) +f_1 \left( qx \right) 
 \right) f_2 \left( {x'} \right) }{ \left( q-1 \right) x}}-{\frac {
 \left( -f_2 \left( {x'} \right) +f_2 \left( q{x'} \right)  \right) 
f_1 \left( x \right) }{ \left( q-1 \right) {x'}}}
\end{equation}
It can be checked that (\ref{Dq}) does not satisfy gauge condition (opposite to its differential limit).
Now substitute $q=e^\hbar$ in (\ref{q-exp}) and expand in powers of $\hbar$:
\begin{equation}
e_q^x={{\mathrm e}^{x}}\left(1-1/4\,{x}^{2}\hbar+ \left( 1/9\,{x}
^{3}+1/32\,{x}^{4} \right) {\hbar}^{2}+ \left( 1/48\,{x}^{2}-
1/16\,{x}^{4}-1/36\,{x}^{5}-{\frac {1}{384}}\,{x}^{6} \right) {\hbar}^{3}+O(\hbar^3)\right)
\end{equation}
We also rewrite (\ref{Ddef}) as $\hbar$-series:
\begin{equation}
D^n=\sum\limits_{0\leq k < \infty}\sigma_k(\partial^0,\partial^1,\dots,\partial^{k+1},x){\hbar}^k,
\end{equation}
where $\sigma_i$ are (to be found) bilinear differential operators with polynomial coefficients. Consider that the
following gauge property 
holds:
\begin{equation}
\sum\limits_{k=0}^{\infty}\sigma_k e_q^x\langle f_1,f_2\rangle {\hbar}^k=e^{2x}\sum\limits_{k=0}^{\infty}\sigma_k \langle f_1,f_2 \rangle {\hbar}^k
\end{equation}
It follows that $\sigma_i$ should satisfy the functional equation:
\begin{equation}
\sigma_i\langle f_1,f_2 \rangle=\sigma_i e^x\langle f_1,f_2 \rangle+\sum\limits_{j=1}^{i}\mathrm{coeff}(\sigma_{i-j}e_q^x\langle f_1,f_2\rangle,{\hbar}^j)
\end{equation}
We choose initial condition:
\begin{equation}
\sigma_0=D(f_1,f_2)=g_{{1}}{g_{2}}_x-g_{{2}}{g_{1}}_x
\end{equation}
Now our coefficients can be written as
\begin{equation}
\sigma_i=\sum\limits_{k=2}^{i+1}a_{i,k}(x)Q_i(\partial), \quad Q_j=g_{{1}}{g_{2}}_{jx}-g_{{2}}{g_{1}}_{jx}
\end{equation}
so we can compare $D^n$ with $\Delta_q$ expansion! The first few coefficients are
\begin{figure}[h!]
\begin{tabular}{r|l}
$\sigma_1$ & $1/4\,{x}^{2}Q_{{2}}$ \\
$\sigma_2$ & ${x}^{2} \left( {\frac {1}{72}}\,xQ_{{2}}+1/24\,{x}^{2}Q_{{3}} \right)$  \\
$\sigma_3$ & ${x}^{3} \left( {\frac {1}{288}}\,{\frac { \left( 5\,{x}^{2}-6 \right) 
Q_{{2}}}{x}}+{\frac {1}{216}}\,{x}^{2}Q_{{3}}+{\frac {1}{192}}\,{x}^{3
}Q_{{4}} \right)$ \\
$\sigma_4$ & ${x}^{4} \left( -{\frac {1}{43200}}\,{\frac { \left( -300+503\,{x}^{2}
 \right) Q_{{2}}}{x}}+ \left( {\frac {23}{3888}}\,{x}^{2}-{\frac {1}{
144}} \right) Q_{{3}}+{\frac {1}{1152}}\,{x}^{3}Q_{{4}}+{\frac {1}{
1920}}\,{x}^{4}Q_{{5}} \right)$ \\
$\sigma_5$ & ${x}^{5} \left( {\frac {1}{259200}}\,{\frac { \left( 540+2431\,{x}^{4}-
2175\,{x}^{2} \right) Q_{{2}}}{{x}^{3}}}+ \left( {\frac {5}{2592}}-{
\frac {173}{48600}}\,{x}^{2} \right) Q_{{3}}+ \left( -{\frac {1}{768}}
\,x+{\frac {47}{41472}}\,{x}^{3} \right) Q_{{4}}+\right.$ \\
& $\left.+{\frac {1}{8640}}\,{x}^{4}Q_{{5}}+{\frac {1}{23040}}\,{x}^{5}Q_{{6}} \right) 
$
\end{tabular}
\end{figure}

\newpage

\section{Gauge operators $D_{(g)}$}
Again, we are dealing with gauge functions analytic in $\hbar$:
\begin{equation}
g(x_1,x_2,\dots,\hbar)=e^{x+\sum\limits_{i=1}^{\infty} S_i(x){\hbar}^i},
\end{equation}
We are going to construct $D_{(g)}^n$ in a perturbative way, using the expansion 
\begin{equation}
D_{(g)}^n=D^n+\sum\limits_{i=1}^{\infty}\sigma_i{\hbar}^i, \sigma_i=\sum\limits_{0 \leq l,m\leq i+1}a_{i}\frac{d^l}{dx^l}\left(f_1\right)\frac{d^m}{dx^m}(f_2)
\end{equation}
($D^n=$ Hirota derivative), s.t. the gauge property (\ref{gp}) is satisfied. For simplicity we take $s=0$.
Note that (\ref{gp}) implies
\begin{equation}
D_{(g)}^n\langle g,g\rangle \equiv 0, \quad \forall n \in \mathbb{Z}
\end{equation}
Choose the first 2 coefficients as
\begin{equation}
D^0_{(g)}\equiv{}D^0=f_1f_2, \quad D_{(g)}^1\equiv{}D^1={f_1}'f_2-f_1{f_2}'
\end{equation}
Then, every $D^n_{(g)}$ is a finite bilinear differential operator, acting on a pair of functions $f_1,f_2$.
Now we are ready to represent the main results, achieved with symbolic computations.

\textbf{Thm.1}
\emph{Let $m$ be even, then we have a symmetric formula:}
\begin{equation}
D_{(g)}^m(f,1)\equiv D_{(g)}^m(1,f)={
\frac {d^{m}}{d{x}^{m}}}f \left( x \right)+\sum _{k=0}^{\frac{1}{2}\,m-1}I_{{\frac{1}{2}\,m-k}} \left( x,\hbar
 \right) {\frac {d^{2\,k+1}}{d{x}^{2\,k+1}}}f \left( x \right) ,
\end{equation}
\emph{where $I_s(x,\hbar)=I_s(g)$ are rational in $g,g',\dots,g^{(m)}$} (note here we equal all arbitrary constants to zero just for simplicity, thus killing all higher order $D$'s in the formula).

For instance,
\begin{equation}
D^2_{(g)}=D^2+I_1(g)(f_1 f_2)',
\end{equation}
where $I_1(g)=\sum I_{1,i}(x){\hbar}^i$. It is easy to see that generating function for $\{I_{1,i}\}$ is a natural logarithm:
\begin{figure}[h!]\centering
\begin{tabular}{c}
$\int I_{1,i} dx$ value \\
\hline \\
$-S_{1;{x}}$ \\
$1/2\,{S_{1;{x}}}^{2}-S_{2;{x}}$ \\
$-1/3\,{S_{1;{x}}}^{3}+S_{2;{x}}S_{1;{x}}-S_{3;{x}}$ \\
$1/4\,{S_{1;{x}}}^{4}+S_{3;{x}}S_{1;{x}}-S_{2;{x}}{S_{1;{x}}}^{2}-S_{4;
{x}}+1/2\,{S_{2;{x}}}^{2}$ \\
$-1/5\,{S_{1;{x}}}^{5}-{S_{2;{x}}}^{2}S_{1;{x}}+S_{4;{x}}S_{1;{x}}-S_{3
;{x}}{S_{1;{x}}}^{2}+S_{2;{x}}{S_{1;{x}}}^{3}+S_{3;{x}}S_{2;{x}}-S_{5;
{x}}$ \\ 
\vdots
\end{tabular}
\end{figure}
\begin{equation}
G(z)=\log \left(\sum\limits_{i=1}^{\infty}t_iz^i\right), S_{j;x} \leftrightarrow \frac{\sum _{i=0}^{\infty }t_{{i}}\frac{i!}{(i-j)!}}{\sum _{i=0}^{\infty }t_{{i}}}
\end{equation}
Then, $I_2(g)$ can be expressed in terms of $I_1(g)$!
\\
\begin{tabular}{c}
$I_{2,i}$ value \\
\hline \\
$I1_{1,x,x}$ \\
${\it I1}_{{2,x,x}}+1/2\,{\it I1}_{{1,x}}{\it I1}_{{1}}$ \\
${\it I1}_{{3,x,x}}+1/2\,{\it I1}_{{2,x}}{\it I1}_{{1}}+1/2\,{\it I1}_{
{1,x}}{\it I1}_{{2}}-{\frac {5}{36}}\,{{\it I1}_{{1}}}^{3}
$ \\
${\it I1}_{{4,x,x}}+1/2\,{\it I1}_{{3,x}}{\it I1}_{{1}}+1/2\,{\it I1}_{
{1,x}}{\it I1}_{{3}}+1/2\,{\it I1}_{{2}}{\it I1}_{{2,x}}-{\frac {5}{12
}}\,{\it I1}_{{2}}{{\it I1}_{{1}}}^{2}
$ \\
${\it I1}_{{5,x,x}}+1/2\,{\it I1}_{{4,x}}{\it I1}_{{1}}+1/2\,{\it I1}_{
{1,x}}{\it I1}_{{4}}+1/2\,{\it I1}_{{2}}{\it I1}_{{3,x}}+1/2\,{\it I1}
_{{3}}{\it I1}_{{2,x}}-{\frac {5}{12}}\,{\it I1}_{{3}}{{\it I1}_{{1}}}
^{2}+{\frac {5}{72}}\,{{\it I1}_{{2}}}^{2}{\it I1}_{{1}}$ \\ 
\vdots
\end{tabular}
\\
Summarizing these facts, we can write the first few $I$'s in closed form:

\begin{tabular}{|l l|}
\hline
$I_{{1}}(g)=$ & $-{\dfrac {g_{{x,x}}}{g_{{x}}}}+{\dfrac {g_{{x}}}{g}}$ \\
$I_{{2}}(g)=$ & $I_{1,x,x}+\frac{3}{2}I_{1,x}^2-5I_1^3$ \\
$I_3(g)=$ & $I_{{1,x,x,x,x}}+5\,I_{{1,x}}I_{{1,x,x}}+10\,I_{{1}}I_{{1,x,x,x}}+15\,I
_{{1}}{I_{{1,x}}}^{2}-65\,{I_{{1}}}^{2}I_{{1,x,x}}-250\,{I_{{1}}}^{3}I
_{{1,x}}+271\,{I_{{1}}}^{5}$ \\
\hline
\end{tabular}

Note that
\begin{equation}
f_1'''f_2-(f_1'f_2')'+f_1f_2'''=\frac{d}{dx}D^2(f_1,f_2)
\end{equation}
Solution for $m\leq 8$ in explicit form:
\begin{equation}
D^4_q=D^4+I_2(g)(f_1f_2)'-\frac{1}{6}I_1(g)(f_1'''f_2-(f_1'f_2')'+f_1f_2''')+R_4(g)(f_1'f_2'),
\end{equation}
\begin{equation}
D^6_q=D^6+I_3(g)(f_1f_2)'+\frac{1}{15}I_2(g)(f_1'''f_2-(f_1'f_2')'+f_1f_2''')+\frac{1}{15}I_1(g)(D^4(f_1,f_2))'+
\end{equation}
\[
+R_6(g)(f_1'f_2')+R_4(g)(-c1(f_1''f_2'')+c2(f_1'f_2'''+f_1'''f_2')),
\]
\begin{equation}
D^8_q=D^8+\left[I_4(g)\frac{d}{dx}D^0+\frac{1}{28}I_3(g)\frac{d}{dx}D^2+\frac{1}{70}I_2(g)\frac{d}{dx}D^4+\frac{1}{28}I_1(g)\frac{d}{dx}D^6\right](f_1,f_2)+R_8D^0(f_1',f_2')
\end{equation}
where $R_4(g)=h^2(-24\,S_{1;{x,x}} \left( S_{1;{x}}S_{1;{x,x}}-S_{2;{x,x}} \right) )+h^3(12\,{S_{2;{x,x}}}^{2}+36\,{S_{1;{x,x}}}^{2}{S_{1;{x}}}^{2}-24\,{S_{1;{
x,x}}}^{2}S_{2;{x}}+24\,S_{1;{x,x}}S_{3;{x,x}}-48\,S_{1;{x,x}}S_{2;{x,
x}}S_{1;{x}}
)+O(\hbar^4)$, and $R_6(g)= c_1(I_1{I_1}_{x,x})_h+c_2(I_1^2{I_1}_{x})_{h,h}$, $c_i$ -- complex constants.

Remarkable fact is that $R_4(g)$ satisfy the following equation:
\begin{equation}
\sum\limits_{i=1}^{\infty} \frac{\partial}{\partial \lambda_i} R_4(g,\lambda_1,\lambda_2,\dots)=I_1(g)\frac{\partial}{\partial h}I_1(g),\quad  \lambda_0=1, \lambda_i=\lambda_{i-1}-\frac{1}{\mathrm{binomial}(i+2,2)},
\end{equation}
We conjecture that similar equalities should hold for any $R_j(g)$.

Now we are ready to represent the general formula.

\textbf{Thm.2} \emph{Gauge operators $D_{(g)}$ are expressed as follows:}
\begin{equation}
\boxed{D_{(g)}^{2k}=D^{2k}+\sum\limits_{1\leq i<k}c_{k-i}I_{k-i}(g)\frac{d}{dx}\left(D^{2i}(f_1,f_2)\right)+{R_{2k}(g)D^0(f_1',f_2')} }
\end{equation}
\begin{equation}
\boxed{D_{(g)}^{2k+1}=D^{2k+1}+\sum\limits_{1\leq i<k}c_{k-i}I_{k-i}(g)\left(D^{2i+1}(f_1',f_2)\right)+{R_{2k+1}(g)D^0(f_1',f_2)}}
\end{equation}
Note here that for odd powers $D^{2k+1}_{(g)}\langle f_1,f_2 \rangle \neq D^{2k+1}_{(g)}\langle f_2,f_1 \rangle$,
which is not true for ordinary $D$'s.

\section{Speculations}

Let $g$ be a series of certain combinatorial nature.
As an example, we consider Hurwitz-Kontsevich genetaring function \cite{3,4}:
%
%
%
\begin{equation}
g=e^{\hbar\hat{W_0}}e^{p_1}
\end{equation}
where $\hat{W_0}$ is the cut-and-join operator:
\begin{equation}
\hat{W_0}=1/2\,\sum _{b=1}^{\infty } \left( \sum _{a=1}^{\infty } \left( a+b
 \right) p_{{a}}p_{{b}}{\it Dp}_{{a+b}}+abp_{{a+b}}{\it Dp}_{{a}}{\it 
Dp}_{{b}} \right), \quad {\it Dp_i}=\frac{\partial}{\partial p_i}
\end{equation}
Now take $\log g=H(\overline{p},q)$, where $H(\overline{p},q)$ -- Hurwitz partition function. 
Its first few terms are:
\\
$$
g_{HK}= \left( 1+1/2\,p_{{2}}{\hbar}+1/2\, \left( 1/4\,{p_{{2}}}^{2}+p_{{3}}+1/2\,{
p_{{1}}}^{2} \right) {h}^{2}+1/6\, \left( 1/2\,p_{{2}}+4\,p_{{4}}+3/4
\,p_{{2}}{p_{{1}}}^{2}+3/2\,p_{{2}}p_{{3}}+1/8\,{p_{{2}}}^{3}+4\,p_{{2
}}p_{{1}} \right) {h}^{3}+1/24\, \left( 4\,{p_{{1}}}^{3}+9\,p_{{3}}+3/
2\,{p_{{2}}}^{2}p_{{3}}+1/2\,{p_{{1}}}^{2}+25\,p_{{5}}+8\,{p_{{2}}}^{2
}p_{{1}}+3\,p_{{3}}{p_{{1}}}^{2}+3/4\,{p_{{1}}}^{4}+13\,{p_{{2}}}^{2}+
27\,p_{{3}}p_{{1}}+3/4\,{p_{{2}}}^{2}{p_{{1}}}^{2}+8\,p_{{2}}p_{{4}}+\right.\right.
$$
\\
$$
+3
\left.\left.\,{p_{{3}}}^{2}+1/16\,{p_{{2}}}^{4} \right) {h}^{4}+{\frac {1}{120}}\,
 \left( 5/4\,p_{{3}}{p_{{2}}}^{3}+40\,p_{{2}}p_{{1}}+15/2\,p_{{3}}{p_{
{1}}}^{2}p_{{2}}+40\,p_{{3}}p_{{4}}+{\frac {125}{2}}\,p_{{5}}p_{{2}}+
10\,{p_{{2}}}^{3}p_{{1}}+{\frac {125}{4}}\,{p_{{2}}}^{3}+20\,{p_{{1}}}
^{2}p_{{4}}+{\frac {15}{8}}\,{p_{{1}}}^{4}p_{{2}}+{\frac {215}{2}}\,p_
{{2}}p_{{3}}p_{{1}}+216\,p_{{6}}+30\,p_{{2}}{p_{{1}}}^{3}+256\,p_{{4}}
p_{{1}}+{\frac {487}{2}}\,p_{{2}}p_{{3}}+\right.\right.
$$
$$
\left.\left.+1/32\,{p_{{2}}}^{5}+{\frac {
495}{4}}\,p_{{2}}{p_{{1}}}^{2}+160\,p_{{4}}+1/2\,p_{{2}}+10\,p_{{4}}{p
_{{2}}}^{2}+5/8\,{p_{{2}}}^{3}{p_{{1}}}^{2}+15/2\,{p_{{3}}}^{2}p_{{2}}
 \right) {h}^{5}+O({\hbar}^6) \right) {{ e}^{p_{{1}}}}
$$
\\
Then we have:
\begin{equation}
H(\overline{p},q), \quad D_{(g)}^2=D^2(f_1,f_2)-\frac{H_{2p_1}}{H_{p_1}}\left({(f_1)}_{p_i}f_2+\delta \cdot f_1{(f_2)}_{p_j}\right),
\end{equation}
where $\delta=0\ \mathrm{or}\ 1$, and the coefficient
\begin{equation}
I_1(g)=-\frac{H_{2p_1}}{H_{p_1}}=\frac{1}{2}\,{h}^{2}+ \left( -\frac{1}{24}-\frac{3}{4}\,p_{{1}} \right) {h}^{4}-\frac{5}{3}\,p_{{2}}{h
}^{5}+\dots
\end{equation}
In these formulas $I_k(g)$ are expressed in terms of logarithmic derivative of Hurwitz-Kontsevich function
and its derivatives.
Now we can check that for the following commutatuor
\begin{equation}
[D_{(g)}^m(f_1,1),D_{(g)}^n(f_1,1)], \quad m,n \ \mathrm{even}
\end{equation}
all coefficients also depend on $\partial\log H$ in a non-trivial way. 
We conjecture that, using this formalism, one can rediscover identities for Hurwitz function.
Also note that if $g$ is a non-trivial tau-function itself, then $D_{(g)}^m(g,g)$ degenerates to some KP-like equation, while $D_{(g)}^m(f_1,f_2)$ corresponds to its minor deformation.




\end{document}